\documentclass[aps,pre,reprint,twocolumn,superscriptaddress]{revtex4-1}
\usepackage{bm}  

\usepackage[]{graphicx}
\usepackage{color}
\usepackage{amssymb,amsmath}

\begin{document}

%\DOIsuffix{theDOIsuffix}

%\Volume{46}
%\Month{01}
%\Year{2007}
%
%
%\pagespan{1}{}
%
%
%\Receiveddate{XXXX}
%\Reviseddate{XXXX}
%\Accepteddate{XXXX}
%\Dateposted{XXXX}

\keywords{Colloids, Yukawa model.}

%% \pretitle{Editor's Choice}

\title[Low-density crystals in charged colloids]{Low-density crystals in charged colloids : Comparison with Yukawa theory}

\author{Ioatzin R\'{i}os de Anda}%
\affiliation{H. H. Wills Physics Laboratory, University of Bristol, Tyndall Avenue, Bristol BS8 1TL, UK}
\affiliation{Bristol Centre for Functional Nanomaterials, Centre for Nanoscience and Quantum Information, Tyndall Avenue, Bristol BS8 1FD, UK}

\author{Antonia Statt}%
\affiliation{Institut f\"ur Physik, Johannes Gutenberg-Universit\"at Mainz,Staudinger Weg 9, 55128 Mainz, Germany}
\affiliation{Graduate School of Excellence Materials Science in Mainz, Staudinger Weg 9, 55128 Mainz, Germany}

\author{Francesco Turci}%
\affiliation{H. H. Wills Physics Laboratory, University of Bristol, Tyndall Avenue, Bristol BS8 1TL, UK}
           
\author{C. Patrick Royall}%
\affiliation{H. H. Wills Physics Laboratory, University of Bristol, Tyndall Avenue, Bristol BS8 1TL, UK}
\affiliation{Centre for Nanoscience and Quantum Information, Tyndall Avenue, Bristol BS8 1FD, UK}
\affiliation{School of Chemistry, University of Bristol, Bristol BS8 1TS, UK}

%\footnote{Corresponding author\quad E-mail:~\textsf{paddy.royall@bristol.ac.uk},            Phone: +00\,999\,999\,999, Fax: +00\,999\,999\,999}}

\begin{abstract}
Charged colloids can behave as Yukawa systems, with similar phase behaviour. Using particle-resolved studies, we consider a system with an unusually long Debye screening length which forms crystals at low colloid volume fraction $\phi \approx 0.01$. We quantitatively compare this system with the Yukawa model and find that its freezing point is compatible with the theoretical prediction but that the crystal polymorph is not always that expected. In particular we find body-centred cubic crystals where face-centred cubic crystals are expected.
\end{abstract}

\maketitle

\section{Introduction}

\begin{figure*}[t]
\includegraphics[width=\linewidth]{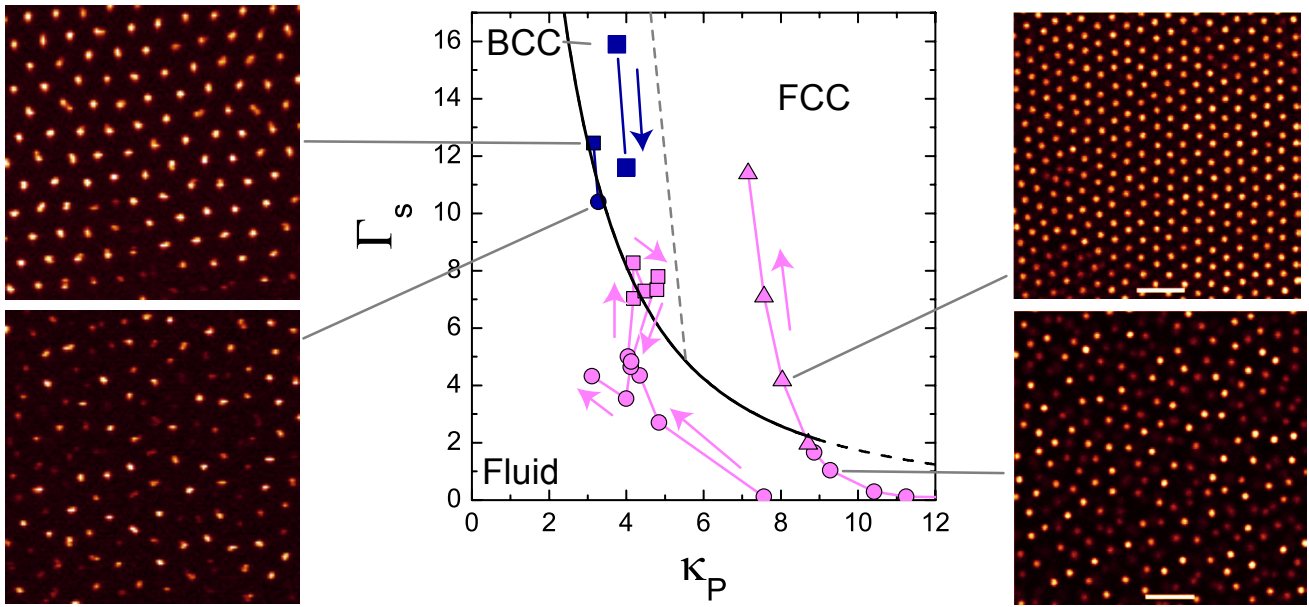}
\caption{Phase diagram of Yukawa systems. The fluid-solid phase boundary (solid line) is the analytic approximation [Eq. \ref{eqGamma}], the dashed line at large $\kappa_p$ denotes that its validity is limited by $\kappa_p<10$. The approximate position of the BCC-FCC crystal boundary is indicated by the dotted line. Narrow coexistence regions are not shown here. Symbols represent various crystallization/melting experiments in colloidal dispersions ~\cite{royall2003,royall2006}(pink) and complex plasmas  ~\cite{khrapak2011}(blue). Squares and bullets indicate, respectively, BCC and fluid phases (as observed), arrows show the direction in which parameters varied during the experiments. Triangles are FCC crystals. Characteristic snapshots of observed fluid and crystalline phases are also shown ~\cite{ivlev}.}
\label{figYukawa}
\end{figure*}

Among the great successes in describing interactions between colloidal particles in suspension is the Derjaguin, Landau, Verwey and Overbeek (DLVO) model ~\cite{verwey1948}. Apart from short ranged interactions, this model treats charged colloids as Yukawa particles. Thus colloids may be interpreted within a framework which encompasses a great many other systems from the mesons which Yukawa was originally interested in ~\cite{yukawanobel} to dusty, or complex plasmas ~\cite{ivlev,loewen2011}. Indeed, as shown in Fig. \ref{figYukawa}, charged colloids and complex plasmas both exhibit the same phase behaviour : under the right conditions both systems exhibit phases characteristic of the Yukawa system, that is to say fluid (F) along with body-centred and face-centred cubic crystals (BCC and FCC respectively).

In the case of colloidal dispersions, there are more components than just the colloids : the particles are immersed in a solvent, the electrostatic charge they carry is balanced by counter-ions (in colloidal dispersions, one typically assumes charge neutrality) and salt ions, not to mention the liquid solvent in which the system is immersed ~\cite{ivlev}. One can proceed to an \emph{effective one-component} system where only the colloids are considered by integrating out the degrees of freedom of the smaller species ~\cite{likos2001}. Here the liquid solvent has no impact on the equilibrium phase behaviour (it acts to damp the dynamics of the colloids, leaving the system as non-inertial on most reasonable timescales)  ~\cite{ivlev}. However the effects of the ions do need to be integrated out, and this can be done using the approach pioneered by Derjagiun, Landau, Verwey and Overbeek ~\cite{verwey1948}. In addition to the effects of the electrostatics, DLVO theory also includes other interactions between the colloids, such as van der Waals forces. In our systems, these are not important, because the colloids are refractive index matched to the solvent which reduces the effects of the van der Waals forces to a fraction to the thermal energy $k_BT$. Any residual van der Waals effects are suppressed by a polymer layer at the surface of the particle. The polymer layer is much thinner ($\lesssim 10 \mu$m) than the particle size ~\cite{bryant2002} and thus the short ranged interactions may be treated as a hard core ~\cite{royall2013myth}.

In colloidal dispersions, the particle concentration is often high enough that steric interactions due to the finite particle size can come into play ~\cite{hynninen2003}, and it is appropriate to include the hard core $u_\mathrm{hc}(r)$. The hard core Yukawa interaction then reads 

\begin{eqnarray}
u(r)&=&u_\mathrm{hc}(r)+ u_{y}(r),
\label{eqU}\\
u_{y}(r)& =& \epsilon_{y}\frac{\exp [-\kappa ( r-\sigma ) ]  }{r/\sigma}.
\label{eqYukawa}
\end{eqnarray}
Here, the potential at contact (when the colloids touch) is given by
\begin{equation}
\beta\epsilon_{y}=\frac{Z^{2}}{(1+\kappa\sigma/2)^{2}}\frac{\lambda_{B}}{\sigma},
\label{eqEpsilonYukawa}
\end{equation}

\noindent where $Z$ is the colloid charge, and the inverse Debye screening length is given by $\kappa$=$\sqrt{4\pi\lambda_{B}\rho_{ion}}$, where
$\rho_{ion}$ is the number density of monovalent ions. The Bjerrum length 

\begin{equation}
\lambda_{B}=\beta e^{2}/(4\pi\epsilon_{0}\epsilon_{r}), 
\label{eqBjerrum}
\end{equation}

\noindent  is the distance at which the interaction energy between two electronic charges is $k_B T$, where $e$ is the electronic charge, $\epsilon_{0}$ the permittivity of free space, and $\epsilon_{r}$ the dielectric constant.

In their investigations of hard core Yukawa phase diagrams, Hynninen and Dijkstra ~\cite{hynninen2003} showed that the effect of the hard core was small in the case that freezing occurred at a colloid volume fraction $\phi \lesssim 0.3$, i.e. that the colloids are typically far enough apart from one another that short-range interactions are irrelevant. Example interaction parameters would be $(\beta \epsilon_y=20,\kappa\sigma=1.0)$. At lower concentrations, therefore, the system can be treated as a point Yukawa system, as can also be the case with complex plasmas ~\cite{ivlev}. A point Yukawa treatment where the hard core is neglected enables arbitrary Yukawa parameters to be represented in a 2d plot  ~\cite{robbins1987,hamaguchi1997}. Here we shall use the Yukawa (screened Coulomb) coupling parameter $\Gamma_s$ and scaled inverse Debye length $\kappa_p$ which are defined below \cite{ivlev}.

In colloidal dispersions, the dominance of such long-ranged interactions means that there is no specific requirement that strongly attractive but short ranged van der Waals interactions be suppressed and thus we are not limited to sterically stabilised, refractive index matched systems. Water and ethanol are popular solvents and silica and polystyrene are popular materials for the colloids. As an aside, this indicates that the melamine particles often used in complex plasma experiments are in principle no different to particles used in colloidal experiments, moreover their size ranges overlap ~\cite{ivlev}.

While the DLVO theory is only valid in the range that linearised Poisson-Boltzmann theory holds (weak electrostatic interactions), higher charging can also be treated with a Yukawa interaction by using a  \emph{renormalised} or \emph{effective} charge that is smaller than the physical charge on the particles ~\cite{alexander1984,trizac2002}. Thus, providing the effective colloid charge can be found, a Yukawa behaviour is recovered. Once the effective charge $Z_\mathrm{eff}$ and Debye length $\kappa^{-1}$ are determined, a number of studies have been made in aqueous based systems finding excellent agreement with the Yukawa model \cite{palberg1999,monovoukis1989,sirota1989,yoshizawa2012}.

Behaviour inconsistent with the DLVO model has also been seen. The DLVO model recasts interactions between multiple components into a one-component Yukawa treatment. Even after accounting for charge renormalisation, anomalous behaviour has been observed. In particular the observation of condensation like behaviour of the colloids into a ``colloidal liquid'' and ``colloidal gas'', with voids appearing in the system, was attributed to ``like charge attraction'' ~\cite{ito1994}. A variety of explanations have been put forward to explain this phenomenon, many arriving at the conclusion that the effective interaction between the colloids was attractive, a surprise for particles with like charge ~\cite{kepler1994,larsen1997,han2003}. However direct measurement with optical tweezers found no evidence of attraction, rather that some earlier measurements (though not the original observation by Ito \emph{et al.} ~\cite{ito1994}) may have been influenced by artefacts ~\cite{baumgartl2005,baumgartl2006}. Among the few theoretical explanations to have withstood the test of time is that of van Roij and coworkers who considered that the entropy of the salt ions might drive phase separation to a colloid-rich and colloid-poor phase ~\cite{vanroij1997,vanroij1998}. Crucially, in van Roij's treatment, at the two-body colloid-colloid level, a repulsion between the particles is maintained : the (repulsive) Yukawa form in Eq. \ref{eqYukawa} is satisfied. The entropic terms driving the phase separation do not feature in the (one-component) DLVO treatment because degrees of freedom of the small ions are \emph{integrated out} and captured in a one-body term. Quantitative agreement with van Roij's predictions was recently found in a phase separating binary system whose behaviour would similarly not be expected in a pure Yukawa picture ~\cite{yoshizawa2012}. Other deviations from DLVO behaviour include ion-colloid decoupling leading to a macroscopic electric field which results in extended sedimentation profiles ~\cite{piazza1993,rasa2004,royall2005s}.

Of particular interest here is a mechanism originally put forward to account for the condensation effects observed by Ito \emph{et al.} ~\cite{ito1994}. While the two-body term between the colloids (Eq. \ref{eqYukawa}) is repulsive, the three body interactions induced between three colloids are \emph{attractive} \cite{russ2002}. Such deviations from two-body behaviour have been observed in experiment in the form of a perturbation to the fluid structure \cite{brunner2002}, and indications of non-spherical interaction potentials in crystals at higher density ~\cite{reinke2007}. Furthermore simulation work indicates that including the three-body terms leads to an increase in the fluid region of the phase diagram at the expense of the BCC phase \cite{hynninen2003,hynninen2004}.

Now the self-dissociation of water means the ionic strength is $\gtrsim 10^{-7}$ Mol and additional contributions such as counter-ions mean that a Debye screening length of $\kappa^{-1} \lesssim 300$ nm is typical in experiment. Since particle resolved studies require colloid sizes of at least a micron, with aqueous solvents it is difficult to reach conditions where the Debye length is comparable to (or greater than) the particle size appropriate for the regime where the system behaves purely as a Yukawa system without significant contribution from the hard core $u_\mathrm{hc}$ and other e.g. van der Waals short range forces. However in solvents such as cycle hexyl bromide of interest here the ionic strength can reach $10^{-12}$ Mol so the Debye length can be sufficient that micron-sized particles are far apart and (point) Yukawa behaviour is found ~\cite{royall2003,royall2006}. In these systems, colloidal crystals of exceptionally low volume fraction, where the inter particle spacing can run to tens of microns have been observed ~\cite{yethiraj2003,royall2003,leunissenThesis}. These systems have also been observed to become dynamically arrested, and to fail to crystallise and thus form a glass at low colloid density ~\cite{klix2010}, as indeed have aqueous systems ~\cite{sirota1989}.

Here we consider the Yukawa parameters associated with such ``low-density crystals''.  Our purpose is to make a quantitative comparison between the low-density crystals and the predictions of Yukawa theory in the form of the equilibrium phase diagram ~\cite{robbins1987,hamaguchi1997}. In particular we map our experimental data to generic Yukawa parameters $(\Gamma_s,\kappa_p)$ and compare the state observed in experiment to the theoretical prediction. We consider the crystal polymorph observed with that predicted. Our analysis indicates that while the Yukawa phase diagram predicts face-centred cubic crystals for some parameters, in our experiments we find body-centred cubic crystals only. This contribution is organised as follows : in Section \ref{sectionMapping} we discuss our mapping procedure and assumptions, in Section \ref{sectionExperimental} we outline our experimental technique. We present our results in Section \ref{sectionResults} and discuss these in section \ref{sectionDiscussion}. 

\section{Mapping to Yukawa theory}
\label{sectionMapping}

The comparison between charged colloids and complex plasmas is illustrated in Fig. \ref{figYukawa}. The main panel in Fig. \ref{figYukawa} is the Yukawa phase diagram in the $\Gamma_s$, $\kappa_p$ plane. Here we show literature data for complex plasmas ~\cite{khrapak2011} and colloids ~\cite{royall2003,royall2006}. We emphasise that the colloidal particles illustrated in Fig. \ref{figYukawa} are two microns in size. They are thus effectively identical to particles used at the smaller end of complex plasma experiments \cite{ivlev}. The key difference is thus the immersing medium, a liquid solvent rather than a plasma.

The freezing line in Yukawa systems is given with reasonable accuracy for $\kappa_p<10$ \cite{ivlev} by

\begin{equation}
\Gamma_s(\kappa_p)=\frac{106}{1+\kappa_p + \frac{1}{2}\kappa_p^2}
\label{eqGamma}
\end{equation}

\noindent where the screened coupling parameter $\Gamma^{(s)}$ is the Yukawa interaction evaluated at the mean inter particle separation $\rho^{-\frac{1}{3}}$, $u_y(\rho^{-\frac{1}{3}})$ where $\rho$ is the particle number density and $\kappa_p$  is a scaled inverse Debye length, given by  $\kappa_{p}$=$\kappa\sigma_{c}\rho^{-\frac{1}{3}}$ \cite{ivlev,robbins1987,hamaguchi1997}.

Here we assume the colloids take their \emph{saturated effective} charge $Z_\mathrm{sat}^\mathrm{eff}$. That is to say, the maximum charge given under charge renormalisation. An approximate value is given by:

\begin{equation}
Z_\mathrm{eff}^\mathrm{sat}=\frac{(2+\kappa\sigma)\sigma}{\lambda_{B}}
\label{eqZ}
\end{equation}

\noindent which represents the effective colloid charge  ~\cite{ivlev}.  Thus the number density of ions can be estimated as the effective charge number per colloidal particle due to the counter ions and that from salt and background ions $\rho_\mathrm{salt}$,

\begin{equation}
\rho_\mathrm{ion}=Z_\mathrm{eff}^\mathrm{sat}\rho + \rho_\mathrm{salt}.  
\label{eqRhoIon}
\end{equation}

\noindent Now although no salt is added, some background ions are present, due for example to solvent self-dissociation. Here we treat this contribution $\rho_\mathrm{salt}$ as a free parameter and determine the scaled screening parameter as outlined above.  We shall see below that assuming agreement with the Yukawa freezing line enables the value of $\rho_\mathrm{salt}=10^{-10}$ $m^{-3}$ which corresponds to 8.3 nMol. Finally, the evaluation of the Yukawa interaction at the mean interparticle separation can be obtained from Eq. \ref{eqYukawa}.

\section{Experimental}
\label{sectionExperimental}

We used poly(methyl methacrylate) (PMMA) colloids sterically stabilized with polyhydroxyl steric acid \cite{bosma2002,ohtsuka2008}. The colloids were labelled with fluorescent rhodamine dye to enable fluorescent imaging and had a diameter of 2000 nm and polydispersity of 5\% as determined with static light scattering. Different volume fractions of the particles were suspended in cyclohexyl bromide (CHB) whose dielectric constant $\epsilon_{r}$ = 7.92 and whose refractive index is closely matched to that of the colloids enabling bulk 3d imaging. A rectangular glass capillary with inner dimensions of 0.10 x 1.00 mm (Vitrocom) was filled with the suspensions and sealed on each end with epoxy glue. The samples were studied with confocal laser scanning microscopy, CLSM (Leica SP5 fitted with a resonant scanner), with 543 nm excitation using a NA 63x oil immersion objective. For qualitative imaging, 2D data set was recorded (512 x 512 pixels), whereas for particle tracking, a full scan of the capillary in the z direction was obtained, providing 3D data sets, where care was taken to ensure the pixel size was the same in the three planes.

\section{Results}
\label{sectionResults}

\subsection{Phase behaviour and comparison with Yukawa theory}

\begin{figure*}[t]
\includegraphics[width=0.68\linewidth]{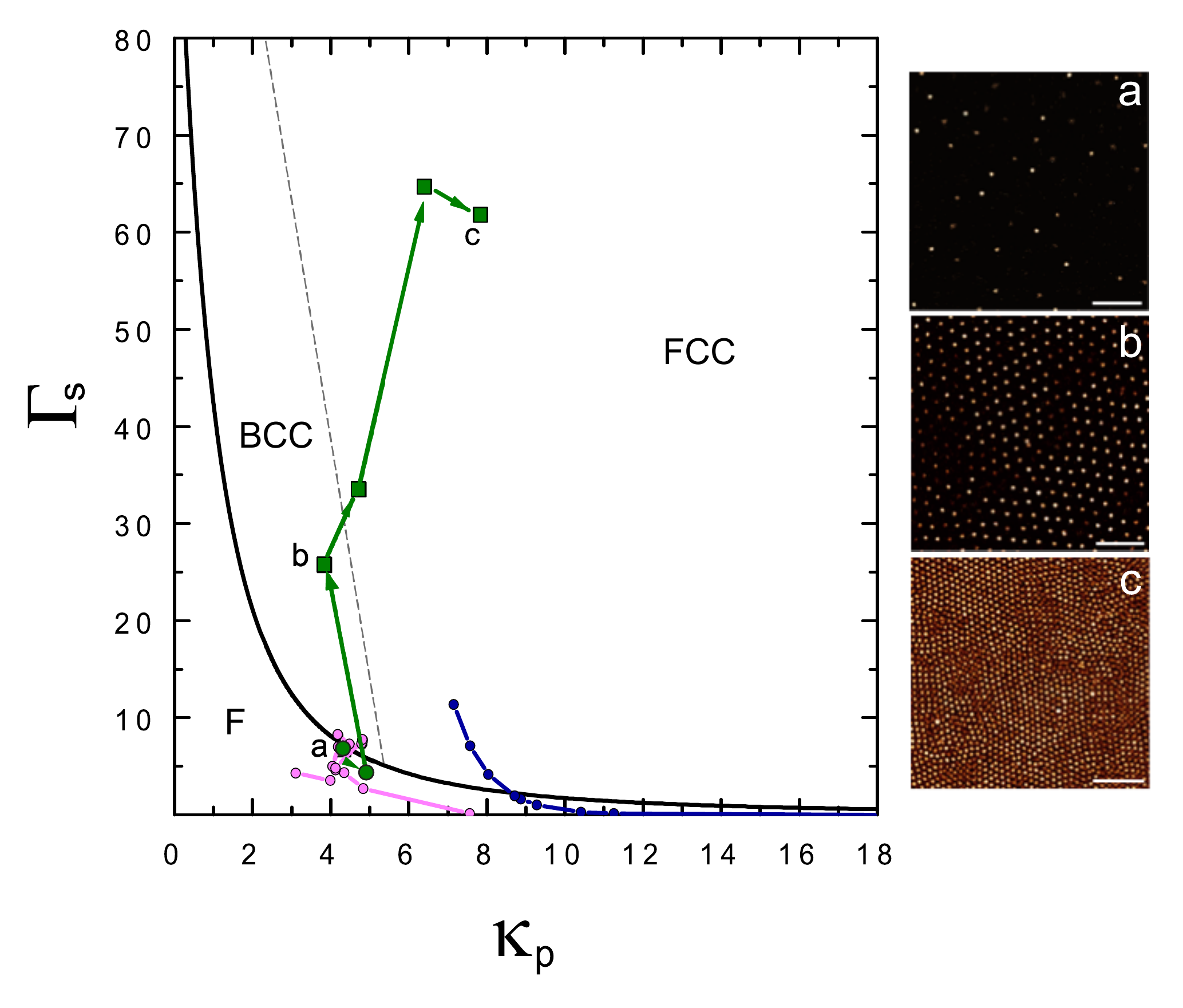}
\caption{Phase diagram of Yukawa system studied in this work in comparison with previous studies  ~\cite{ivlev}.
Snapshots in (a) (b) (c) correspond to volume fraction $\phi$ of 0.0055, 0.02 and 0.23 respectively and the state points in $(\Gamma_s,\kappa_p)$ representation indicated in the main panel. Lines are as in Fig. \ref{figYukawa} : thick line is the freezing line [Eq. \ref{eqGamma}] and the thin dashed line approximately describes the BCC-FCC transition. Arrows denote increasing volume fraction. Scale bars = 25 $\mu$m.
}
\label{figYukawaIoatzin}
\end{figure*}

We show our comparison with Yukawa theory in Fig. \ref{figYukawaIoatzin} for the low-density crystals formed in this work. Compared to previous work where the experiments were mapped to Yukawa parameters ($\Gamma_s,\kappa_p$), we access a new region of the phase diagram. We fit the ionic strength such that our lowest concentrations $\phi$= 0.0055 and 0.01, identified as fluids are consistent with the Yukawa prediction. We note that the path the state points take in the ($\Gamma_s,\kappa_p$) space is not a straight line or even a smooth curve, nor even montonic. This is due to competing effects. Regadring the ionic strength, the added salt $\rho_\mathrm{salt}$ in Eq. \ref{eqRhoIon} is comparable to or smaller than the counter ion contribution $Z_\mathrm{eff}^\mathrm{sat}\rho$ which means that the Debye length $\kappa^{-1}$ drops as the volume fraction is increased so the screening is stronger. However the increase in $\phi$ means the range at which the screened coupling parameter $\Gamma_s$ is evaluated, $\rho^{-\frac{1}{3}}$, drops because the particles are (on average) closer together. The former acts to reduce $\Gamma_s$, the latter to increase it. We note that the highest three densities ($\kappa_p$) are predicted to the FCC. We describe how we determined the crystal structure in the following section.

We are now able to quantify the Debye screening length $\kappa^{-1}= 1.9 \mu$m and contact potential $\beta \epsilon_y=1110$  around freezing. This corresponds to a effective colloid $Z_\mathrm{eff}^\mathrm{sat}=850$. In particular the Debye length is  bigger, compared to aqueous systems and indeed to previous particle-resolved studies where it was around 0.200-1 $\mu$m  ~\cite{royall2003,royall2006}.

\subsection{Identification of crystal structures at low packing fractions}

\begin{figure*}[t!]
\begin{center}
\includegraphics[width=0.8\textwidth]{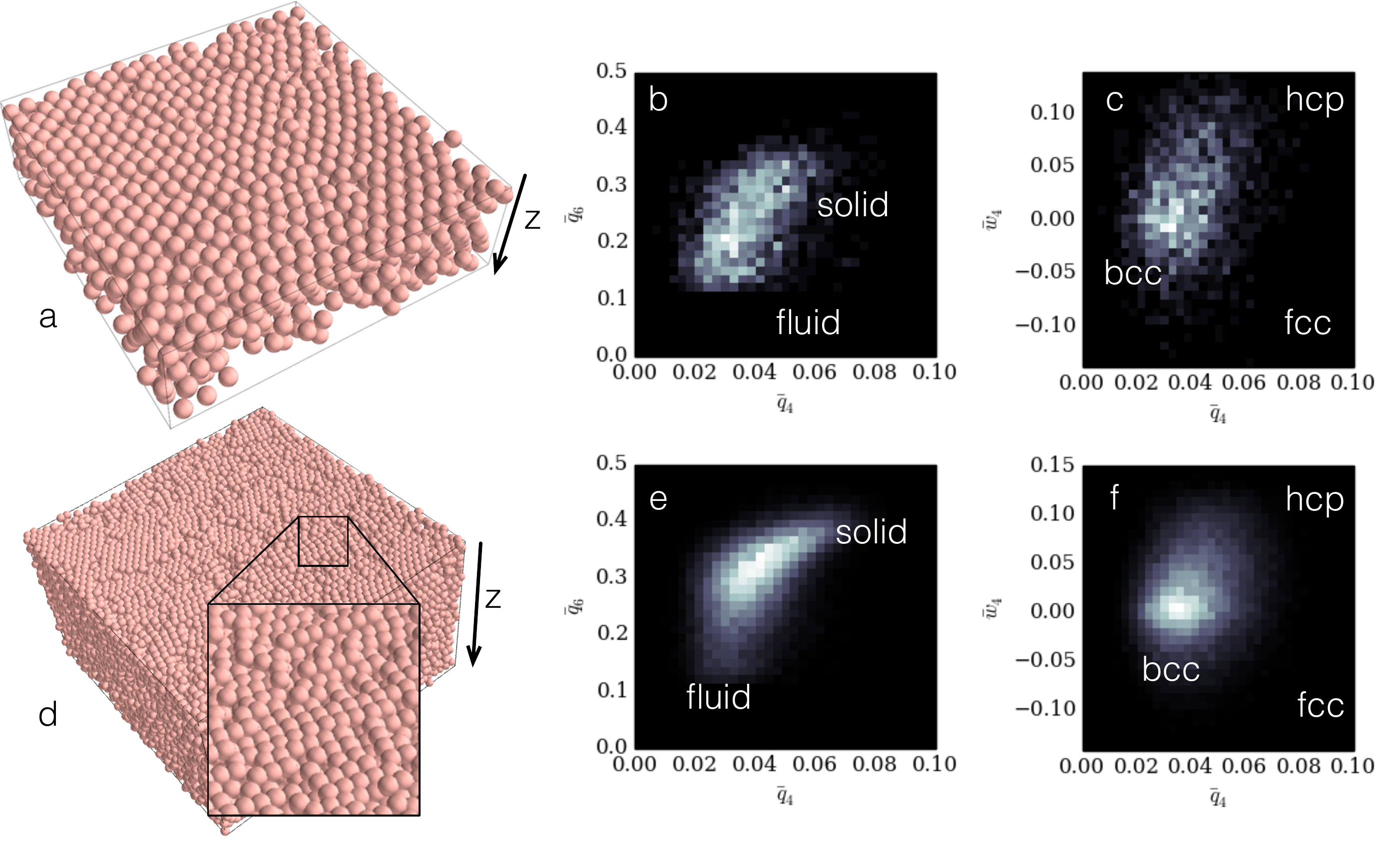}
\end{center}
\caption{
(a,d) Three-dimensional rendering of the dense BCC crystal seen from the bottom of the sample for (a) volume fraction $\phi=0.015$  [corresponding to point b in Fig. \ref{figYukawaIoatzin}] and (d) volume fraction $\phi=0.16$ [point c in Fig. \ref{figYukawaIoatzin}], with arbitrarily scaled particles radii for better visualisation. In panel (d), a close-up of the ordered surface is highlighted. (b-c, e-f) Corresponding local bond order parameter diagrams for the bcc phase of the two considered volume fractions: in panel (b) the distribution $P(\bar{q}_4,\bar{q}_6)$ is centred at moderate values corresponding to a solid region whose nature is determined via the  $P(\bar{q}_4,\bar{w}_4)$ distribution to be of BCC nature. The same procedure is applied in panels (e,f) for the higher volume fraction, showing a higher degree of order, particularly in the $P(\bar{q}_4,\bar{q}_6)$ distribution.}
\label{figBcc}
\end{figure*}

For the analysis of the crystal structures particle tracking of the 3d data sets based on \cite{dunleavy2015} was performed. Image manipulation techniques are used in order to enhance the contrast and remove the noise from the experimental datasets. The positions of the particles are identified through a maximization of the overlap between seeded local gaussian kernels and the intensity profile of the particles. For this algorithm, an estimation of the size of one particle is needed as input parameter, but no initial assumptions on the volume fraction are required. The resulting coordinates are obtained with a resolution of 1 pixel (200 nm) in each spatial direction.

The coordinates extracted via particle tracking are used in order to perform an analysis of the crystalline phases formed before the sedimentation of the sample. For this purpose, we employed the Steinhardt local rotational invariants, also known as bond-orientational order parameters. These discriminate between different possible crystal structures on the basis of spherical harmonics and have also been used for the detection of order in complex plasmas \cite{steinhardt1983,khrapak2011}. In particular, we consider the locally averaged order parameters $\bar{q}_4, \bar{w}_4,\bar{q}_6$ for square and hexagonal order, where the local average allows to take into account the effect of second nearest neighbours and to more sharply distinguish between different arrangements (see \cite{lechner2008} for a detailed discussion on the technique).

For each particle $i$, we perform a parameter-free detection of the nearest neighbours via a Voronoi tessellation of the sample volume. This provides a list of the $N_b (i)$ nearest neighbours over which the local order parameters are calculated:
\begin{equation}
q_{lm}(i)=\frac{1}{N_{b}(i)}\sum_{j=1}^{N_{b}(i)}Y_{lm}(\bm{r}_{ij}) \;,
\end{equation}
where $Y_{lm}(\bm{r}_{ij})$ are the spherical harmonics. Summing over the list  of $\tilde{N}_b(i)$ particles identified by the neighbours and the particle $i$ itself one obtains the locally averaged order parameters
\begin{equation}
\bar{q}_{lm}(i)=\frac{1}{\tilde{N}_{b}(i)}\sum_{k=0}^{\tilde{N}_{b}(i)}q_{lm}(k),
\end{equation}
and summing over all the harmonics we finally get
\begin{equation}
\bar{q}_{l}(i)=\sqrt{\frac{4\pi}{2l+1}\sum_{m=-l}^{l}|\bar{q}_{lm}(i)|^{2}} \;.
\end{equation}
\begin{equation}
\bar{w}_l(i)=\dfrac{\sum\limits_{m_1 +m_2 + m_3=0}  \left(\begin{array}{ccc}l & l & l \\m_1 & m_2 & m_3\end{array}\right) \bar{q}_{lm_1}(i)\bar{q}_{lm_2}(i)\bar{q}_{lm_3}(i) }{\left(\sum_{m=-l}^{l}|\bar{q}_{lm}(i)|^{2}\right)^{3/2}} 
\end{equation}
where the term in brackets is the Wigner symbol.

We focus our analysis on a bulk region discarding top, bottom and lateral edges for a thickness of about 2.5 $\mu$m. In the lowest density samples, we can discriminate between a fluid and a solid phase, where the fluid presents the characteristics of a layered liquid along the vertical $z$ dimension. In Fig.~\ref{figBcc} we show the results of the local order analysis for a low density system (top row, volume fraction $\phi=0.015$) and a dense sample (bottom row,  $\phi=0.16$). The very limited range of the $\bar{q}_4$ order parameter allows us to discard the hypothesis of a face centred cubic crystal. In order to asses the nature of the solid phase, we use an additional order parameter $\bar{w}_4$, particularly suitable for the distinction of hexagonal close packed structures (HCP) and FCC from BCC: in Fig.~\ref{figBcc}(b) we show that no peak is detected in the HCP or FCC region, leading to the identification of the solid phase as a BCC phase. We see that the state point at $\phi=0.16$ [(c) in Fig. \ref{figYukawaIoatzin}] is identified as BCC while the theory predicts that it should be FCC. Indeed we analysed all crystalline systems and found only BCC crystals. We speculate on the causes of this discrepancy below.

\section{Discussion}
\label{sectionDiscussion}

Our work extends the range of Yukawa parameters ($\Gamma_s,\kappa_p$) for particle-resolved studies of colloidal systems. We have shown that the formation of low-density crystals in these systems \cite{yethiraj2003,royall2003,leunissenThesis} is compatible with Yukawa theory, and that Debye screening lengths can run to many microns. The screened coupling parameter is found to be $\Gamma_s=67$, the largest value obtained previously was $\Gamma_s=16$. We further use our parameterisation to estimate the lowest freezing density attainable in the case of no salt ($\rho_\mathrm{salt}\rightarrow0$). This turns out to be around $\phi= 0.0004$ for our system and a Debye length of some $14$ $\mu$m. In principle, therefore it might be possible to produce crystals of very much lower density still.

We now consider possible reasons for the discrepancy of the polymorph we identified. One possibility might be three-body and higher order effects, which could be significant for such long-range interaction potentials. However as we have noted above simulations predict the opposite behaviour, of a suppression of the BCC phase relative to the two-body Yukawa system and the formation of FCC instead ~\cite{hynninen2003,hynninen2004}. Our primary speculation on the other hand is that the system has not yet fully equilibrated. Indeed in some samples, the region close to the wall appearing more crystalline, indicating a possible crystallisation front beginning at the wall of the same cell as has a been observed in the hard sphere suspensions ~\cite{sandomirski2011}. While homogenous crystallisation for our parameters has yet to be studied in detail, we note that in Yukawa systems, although FCC is the favoured polymorph, BCC can form first in the Ostwald rule of stages and indeed polymorph selection can even proceed through the metastable hexagonal close-packed polymorph ~\cite{desgranges2007}. We believe that the same could occur here, although we emphasise that the effect of the wall will be profound in the crystallisation mechanism and thus the homogenous crystallisation studies of Desgranges and Delhomme ~\cite{desgranges2007} may not hold for our case.

However we note that our system is confined in a capillary of height 100 $\mu$m. Confinement has been considered in comparable systems up to four layers ~\cite{oguz2012}. There some preference was found for square symmetry which might lead to BCC being favoured in our larger system. Thus it would be most interesting in the future to investigate the crystallisation kinetics to see if the system indeed showed signs of approaching its FCC bulk equilibrium state. However we emphasise that these systems are stable for around two days ~\cite{royall2003} which does place some limits on the experimental time window. While no change is observed on the timescale of two days (here limit our experiments to four hours), after a week the system was found to behave as of the contact potential $\beta \epsilon_y$ has dropped along with the Debye length. This behaviour is inferred from the shift in the freezing boundary to higher colloid volume fraction for older samples ~\cite{royall2003,royall2006}. Although desirable, an in-depth study of these ageing phenomena has yet to be carried out. Possible sources of the change in the system over time include ion dissolution from the glass capillary in which the system is confined.

\subsection*{Acknowledgements}
The authors would like to thank Alfons van Blaaderen, Marjolein Dijkstra, Bob Evans, Antti-Pekka Hynninen, Alexei Ivlev, Mirjam Leunissen and Hartmut L\"{o}wen for many helpful discussions.  Andrew Dunleavy is particularly thanked for the generous provision of the particle tracking code. CPR would like to acknowledge the Royal Society for financial support. FT and CPR acknowledge the European Research Council under the FP7 / ERC Grant agreement n$^\circ$ 617266 ``NANOPRS''. IRdA was supported by a doctoral scholarship from CONACyT. AS acknowledges financial support by the Deutsche Forschungsgemeinschaft (grant No. VI237/4-3)

%\bibliographystyle{unsrt}
%\bibliography{hacienda}

\end{document}